\documentclass[10pt,letterpaper]{article}

\usepackage[utf8]{inputenc}

\usepackage{textcomp,marvosym}

\usepackage{fixltx2e}

\usepackage{amsmath,amssymb}

\usepackage{cite}

\usepackage{nameref,hyperref}

\usepackage[aboveskip=1pt,labelfont=bf,labelsep=period,justification=raggedright,singlelinecheck=off]{caption}

\usepackage{algorithm,algorithmic}

\makeatletter
\renewcommand{\@biblabel}[1]{\quad#1.}
\makeatother

\date{}

\usepackage{lastpage,fancyhdr,graphicx}
\usepackage{epstopdf}

\usepackage[normalem]{ulem}
\usepackage{color}
\newcommand{\rem}[1]{}
\newcommand{\add}[1]{#1}

\begin{document}
\vspace*{0.35in}

\begin{flushleft}
{\Large
\textbf\newline{\textsl{marathon}: An open source software library for the analysis of Markov-Chain Monte Carlo algorithms}
}
\newline
\\
Steffen Rechner\textsuperscript{1,*},
Annabell Berger\textsuperscript{1,2}
\\
\bigskip
\bf{1} Institute of Computer Science, Martin Luther University Halle-Wittenberg, Halle (Saale), Germany
\\
\bf{2} German Centre of Integrative Biodiversity Research (idiv) Halle-Jena-Leipzig, Leipzig, Germany
\\
\bigskip

%
%





* steffen.rechner@informatik.uni-halle.de (SR)

\end{flushleft}

A version of this article has been published in PLOS ONE, 2016.

\section*{Abstract}
We present the software library \textsl{marathon}, which is designed to support the analysis of sampling algorithms that are based on the Markov-Chain Monte Carlo principle. 
The main application of this library is the computation of properties of so-called state graphs, which represent the structure of Markov chains. We demonstrate applications and the usefulness of \textsl{marathon} by investigating the quality of several bounding methods on four well-known Markov chains for sampling perfect matchings and bipartite graphs. In a set of experiments, we compute the total mixing time and several of its bounds for a large number of input instances. We find that the upper bound gained by the famous canonical path method is often several magnitudes larger than the total mixing time and deteriorates with growing input size. In contrast, the spectral bound is found to be a precise approximation of the total mixing time.


\section*{Introduction}\label{Section:Introduction}

The task of random sampling is to return a randomly selected object from a typically large set of objects according to a specified probability distribution. Such tasks often arise in practical applications like network analysis, where properties of a certain network of interest are to be compared with those of a random null model network~\cite{Kashtan2005, gotelliNicholas}. Another application is approximate counting of combinatorial objects. While this is typically a hard problem, the number of solutions of a \emph{self-reducible} problem, however, can be approximated in polynomial time using randomly sampled objects \cite{Jerrum1986169}. 
We consider sampling methods which follow the so-called \emph{Markov-Chain Monte Carlo} (MCMC) approach. A MCMC algorithm can be considered as a \emph{random walk} on a directed graph where the vertices of this graph correspond to the set of objects to be sampled from. Starting at an arbitrary vertex, we go to a randomly chosen adjacent vertex and continue our random walk from this vertex; after some time, we stop the random walk. The object representing the final vertex is returned as a random sample. An overview of the surprisingly versatile applications of the MCMC approach was given by Diaconis \cite{diaconis2009markov}.

\paragraph*{Motivation}
In general, an infinite number of steps will lead to a truly random sample. But what happens if we stop the random walk after a finite number of steps? The number of steps which are necessary to sample from a probability distribution which is \emph{close} to the desired distribution is known as the \emph{total mixing time} of a Markov chain, and is of central interest for the applicability of an MCMC algorithm. There are several methods which are able to gain upper bounds on the total mixing time like the \emph{canonical path} method. Sinclair \cite{Sinclair92} gives an excellent overview about different bounding techniques.
Sometimes, such bounding methods can successfully be applied to Markov Chains to gain upper bounds on their total mixing times. When applied successfully, such upper bounds most often are high-degree polynomials on the size of the input. This is far too large to be applicable in practice, where more than a linear number of steps is infeasible for large input size. In fact, software tools like \textsl{mfinder} \cite{Kashtan2005}, which can be used for motif search in large networks, use a MCMC sampling approach with a linear number of steps as default for generation of null models.  This leads to the problem that the sampling result might not be as random as desired, resulting in a non-optimal or even incorrect behaviour of the application. 

On the other hand, it may be the case that upper bounds on the total mixing time are just too pessimistic.
From a purely theoretical perspective it is often already a breakthrough to establish that a given Markov chain is rapidly mixing, that is, to establish a polynomial mixing time bound.  Since the bounding techniques in Markov chain analysis are often fairly general and worst-case instances in terms of total mixing time are not known, it is not clear, whether the upper bounds gained by such methods are tight for some worst-case instance. However, for practical applicability, it is of eminent importance to find as sharp bounds as possible. Up to now, there is very little knowledge about the gap between the so established upper bounds and tight bounds for actual worst-case instances. Probably most researches in this field will suspect that a considerable real gap exists, but for specific Markov chains it is in general completely open how many orders of magnitude this gap may be large. By this reason, we believe that the true total mixing time might be much lower than proven by theoretical methods. 
Therefore, our working questions are: a) Is the total mixing time really as large as the bounding methods propose, or is it possible that the bounding methods are just not precise enough to capture the real total mixing time? b) Which bounding method has the best potential and could lead to better results when further information about the structure of state graphs is given?

\paragraph{Contribution}
In trying to answer these questions, we developed the \textsl{C++} library \textsl{marathon}, which has been designed to compute several structural properties of Markov chains and its corresponding state graphs. The library offers the following features:
\begin{itemize}
\item Construction of state graphs from arbitrary input instances for a set of user-definable transition rules.
\item Computation of structural properties of state graphs.
\begin{itemize}
	\item Computation of the total mixing time for arbitrary $\epsilon$.
	\item Computation of the congestion bound gained by the canonical path method with an arbitrary, user-definable, path construction scheme.
	\item Computation of the upper and lower spectral bound.
	\item Computation of network properties like diameter and average path length.
\end{itemize}
\end{itemize}
We built this library to easily add\rem{additional} algorithms for network analysis. We expect that our library could be helpful both for theoretical scientists as well as for practitioners, who implement MCMC algorithms and have to choose an appropriate number of steps. 

As a demonstration on what kind of research can be done with the help of \textsl{marathon}, we analysed structural properties and investigated the quality of the total mixing time bounds on the example of four well-known Markov chains. From our experiments, we gained several insights.
\begin{itemize}
\item We found that the congestion bound is multiple times larger than the corresponding total mixing time for almost all instances. In addition, the quality of the congestion bound deteriorates with growing input size. This indicates that the congestion bounds are bad approximations of the total mixing time, and, since this bound is often used for theoretical analysis, that the high-degree polynomial bounds from theory may be too pessimistic.
\item In contrast, the upper spectral bound is close to the total mixing time for all observations. Even if its quality also deteriorates with growing instance size, the spectral bounds keeps close to the total mixing time much longer than the congestion bound. 
\item The lower spectral bound can be used as a very good approximation for all input instances we investigated in this work. The data suggests an almost linear relationship between the lower spectral bound and total mixing time. 
\item We figured out several new structural insights about the Markov chains we investigated. For example, we found that the the second largest eigenvalue in magnitude is almost always positive. The total mixing time of instances of the same input size can be very diverse and depends on the vertex degree in the state graphs. This shows that the experimental approach provided by our library may lead to structural insights which may be useful to get new intuitions for developing more precise bounding techniques.
\end{itemize} 
From our experiments we can conclude that the canonical path method will not lead to tight bounds of the total mixing time, even when further information about a state graphs structure is included. Instead, developing new methods based on the spectral gap seems to be more promising.

\paragraph{Structure}
The remainder of this article is structured as follows: In the next section we give a brief introduction into the theory of Markov chains, in particular, in the concepts of total mixing time and its bounds. Thereafter, we present the main features of the \textsl{marathon} library. In the second part of this article, we demonstrate the applicability of the library in a set of experiments, while trying to answer the questions described above. The results of these experiments are shown and discussed in the final section.

\section*{Methods and Materials}
An important tool for understanding MCMC based sampling processes are Markov chains. To make the following sections more understandable, we will briefly introduce the most important concepts of the theory of Markov chains. For a less steep introduction into the topic, we recommend the lecture notes of Sinclair \cite{SinclairLectures}, the textbook by Levin, Peres and Wilmer \cite{LevinPeresWilmer2006}, and the survey of Lovász \cite{Lovasz93randomwalks}.

\subsection*{Theory of Markov Chains}\label{subsec:theory}

A Markov chain can be seen as a random walk on a set $\Omega$ of combinatorial objects, the so-called \emph{states}. Two states $x,y \in \Omega$ can be connected via a \emph{transition arc} $(x,y) \in \Psi$, when $x$ can be transformed into $y$ via small local changes. This definition induces a so-called \emph{state graph} $\Gamma = (\Omega, \Psi)$, representing the objects and their adjacencies. We define for each pair of states $x,y \in \Omega$ the so-called \emph{proposal probability}~$\kappa\colon \Omega \times \Omega \to [0,1]$ and a \emph{weight function} $w\colon \Omega \to \mathbb{R}^+$. A step from $x$ to $y$ in a random walk is done with \emph{transition probability} $P(x,y) = \kappa(x,y) \cdot \min(1, w(y)/ w(x) ).$
The matrix~$P = (P(x,y)_{x,y\in \Omega})$ is called \emph{transition matrix}.

Algorithm~\ref{alg:randomwalk} shows the classical random walk based sampling method. A random state $b \in \Omega$ is returned, sampled from a probability distribution $p^{(t)}_a$ which describes a state's probability after~$t$ steps when starting at state~$a \in \Omega$. The \emph{fundamental theorem for Markov chains} (see for example \cite{LevinPeresWilmer2006}) says that the distribution $p^{(t)}_a$ converges for~$t \to \infty$ to the unique \emph{stationary distribution}~$\pi$ with $\pi(y) = \frac{w(y)}{\sum_{x \in \Omega}w(x)}$ for all $y \in \Omega$ and initial states $a \in \Omega$, if the state graph $\Gamma$ is non-bipartite, connected and \emph{reversible} with respect to~$\pi$, i.e. $\pi(x)P(x,y)=\pi(y)P(y,x)$ for each pair of states $x,y\in \Omega$. Markov chains which fulfil these properties are called \emph{ergodic}.
\begin{algorithm}[H]
    \caption{Random Walk} \label{alg:randomwalk}
    \begin{algorithmic}[1]
      \REQUIRE~State~$a \in \Omega$, number~$t$ of steps.
      \ENSURE~State~$b \in \Omega$, according to probability distribution $p_a^{(t)}$.
      \STATE $x \gets a$
      \FOR{$i\gets 1$ \TO $t$}
         \STATE \textbf{Neighbour selection}: Pick a neighbour~$y$ of~$x$ with probability~$\kappa(x,y).$
        	\STATE \textbf{Metropolis rule:} $x \leftarrow y$ with probability~$\min \left( 1, \frac{w(y)}{w(x)} \right).$
      \ENDFOR
      \RETURN $x$
      \end{algorithmic}
      \label{alg:randomwalk}
  \end{algorithm}
  
\paragraph{Total Mixing Time}
The \emph{total variation distance} measures the distance between two probability distributions $\mu$ and $\eta$:
\begin{equation}
|| \mu - \eta || := \frac{1}{2}\sum_{x \in \Omega} | \mu(x) - \eta(x)|.
\end{equation}
We define $\tau_a(\epsilon) := \min \{ t \in \mathbb{N} \colon  || p_a^{(t)} - \pi || \leq \epsilon\}$ as the minimal number of steps a random walk has to take to reach a distribution which is close to $\epsilon$ to its stationary distribution when starting at state $a\in \Omega$. 
The \emph{total mixing time} $\tau(\epsilon)$ of a Markov chain is defined as \begin{equation}
\tau(\epsilon) := \max_{a \in \Omega} \{ \tau_a(\epsilon) \}.
\end{equation}
A Markov chain is known as \emph{rapidly mixing}, if $\tau(\epsilon)$ can be bounded from above by a polynomial which depends on the input size~$n$ and the parameter~$\epsilon^{-1}$. 
The total mixing time can be bounded by several techniques. We briefly present two widely used bounding methods which we investigate in the remainder of this article.

\paragraph{Spectral Bound} Let~$1 = \lambda_1 > \lambda_2 \geq \ldots \geq \lambda_{|\Omega|} > -1$ be the real eigenvalues of the transition matrix~$P$ and let $\lambda_{\max}$ be defined as $\lambda_{\max} = \max \{ |\lambda_2|, |\lambda_{\Omega}| \}$. The total mixing time can be bounded by the lower and upper \emph{spectral bound} \cite{Sinclair92}, i.e.,
\begin{eqnarray}
\tau(\epsilon) &\leq& \left(1-\lambda_{\max}\right)^{-1} \cdot \left( \ln(\epsilon^{-1}) + \ln(\pi_{\min}^{-1}) \right)\label{eqn:upper_spectral_bound}\\
\tau(\epsilon) &\geq& \frac{1}{2}\lambda_{\max}(1-\lambda_{\max})^{-1}\ln(2\epsilon)^{-1}\label{eqn:lower_spectral_bound},
\end{eqnarray}
where $\pi_{min}$ denotes the smallest component of~$\pi.$ 
The transition matrix~$P$ is not necessarily symmetric. However, it can be transformed into a symmetrical matrix:
\begin{equation}
P_{sym} = D^{-1/2} \cdot P \cdot D^{1/2}, \label{eqn:symmetrical}
\end{equation}
where $D$ is the $|\Omega| \times |\Omega|$ diagonal matrix with the components of $\pi$ on its main diagonal. This transformation relies on the reversibility of a Markov chain. It is a classical result that $P_{sym}$ has the same eigenvalues as $P$.

\paragraph{Congestion Bound} Sinclair's \emph{multi commodity flow method} \cite{Sinclair92} is often used for bounding the total mixing time. Let~$\mathcal{P} = \bigcup_{x\not=y} \mathcal{P}_{xy}$ be a family of simple paths in~$\Gamma$, each~$\mathcal{P}_{xy}$ consisting of simple paths between~$x$ and~$y \in \Omega$. The \emph{maximum load congestion} $\rho$, with respect to~$\mathcal{P}$, is then defined as 
\begin{equation}
\rho(\mathcal{P}) = \max_{(u,v) \in \Psi} \frac{\sum_{p \in \mathcal{P} \colon (u,v) \in p} \pi(x)\pi(y)|p|}{\pi(u) P(u,v)}\label{eqn:maximumLoading}.
\end{equation}
For any system of paths~$\mathcal{P}$ the total mixing time of a reversible Markov chain 
can be bounded by the \emph{congestion bound}, i.e.,
\begin{eqnarray}
\tau(\epsilon) &\leq& \rho(\mathcal{P}) \cdot \left( \ln(\epsilon^{-1}) + \ln(\pi_{\min}^{-1}) \right)\label{eqn:flow2}.
\end{eqnarray}
The quality of the congestion bounds depend on the path construction scheme $\mathcal{P}$. The congestion bound is often used to gain theoretical bounds on the total mixing time.

\subsection*{The marathon library}
\label{subsection:software}

In this section, we introduce the main features of the \textsl{marathon} library. Our source code is published under MIT licence and freely available at\\ \url{https://github.com/srechner/marathon}. To install, just follow the instruction manual at github. Several example programs are available.

The \textsl{marathon} library is designed for the study of structural properties of Markov chains, respectively their corresponding state graphs. One of its current main applications is the computation of the total mixing time and several of its bounds. The suggested way to use our library is to implement the transition rules of a Markov chain and conduct some experiments to quickly learn some properties which are typically hard to find in theory. In particular, we allow the computation of the eigenvalue $\lambda_{\max}$ and the application of the canonical path method to compute the congestion bound with some path congestion scheme. This way, one can quickly evaluate whether a scheme captures the total mixing time closely or if there is a noticeable gap. 
The \textsl{marathon} library has been designed  with two main goals in mind: Performance and Extensibility. Aiming for the first goal, we use the \textsl{C++} programming language, so that various highly efficient third party libraries become available. In particular, we use \textsl{CUDA} \cite{cudaToolkit} and \textsl{OpenMP} \cite{OpenMP} to accelerate compute-intense tasks with the use of multi-core processors and highly parallel graphic processing units.
To achieve the second goal, we designed our library for easy extensibility. Our representation of the state graphs is versatile enough to allow the integration of arbitrary graph algorithms. Moreover, we provide interfaces to simplify the programming effort and to hide complexity from the user. For example, adding a new Markov chain is done in a few steps:
\begin{enumerate}
\item Implement a data structure which represents a state, here simply called \texttt{State}.
\item Create a class which implements the \texttt{MarkovChain} interface, using \texttt{State} as a template parameter. The following virtual methods have to be overridden:
\begin{itemize}
\item The method \texttt{MarkovChain::arbitraryState}, constructs an arbitrary state from a given input string (e.g. construct a bipartite graph from a given bipartite degree sequence).
\item The method \texttt{MarkovChain::neighbours}, takes an object $u$ of type \texttt{State} and creates a list of adjacent states $N_u$ whose elements are the result of all valid random choices within $u$ plus their corresponding proposal probability $\kappa$, according to the set of transition rules.
\end{itemize}
\end{enumerate}
In many cases this can be done within about 200 lines of code with a high amount of re-usability. 
Example chains have been included within the library. We will now give a detailed description of the algorithmic ideas behind the most important methods available in the library.

\paragraph{Construction of the State Graph}

Given an instance representation (e.g. a degree sequence), the first step of the analysis is the construction of the state graph for the given instance. The purpose of this step is to gain a sparse representation of the state graph $\Gamma=(\Omega, \Psi)$. 
Initializing $\Omega$ with an arbitrary state generated by the \texttt{MarkovChain::arbitraryState} method, we perform a full graph scan to iteratively enumerate the set of states and transition arcs. In each step, we select a state $u \in \Omega$ and enumerate all its adjacent states~$N_u$ via the \texttt{MarkovChain::neighbours} method. We add a transition arc~$(u,v)$ with proposal probability $\kappa(u,v)$ to $\Psi$ for each adjacent state $v \in N_u$ and add $v$ to~$\Omega$, if not already included.  At the end of this step, we have a complete list $\Omega$ of states and an adjacency list representation of the transition arcs~$\Psi$. 
The probability of each transition $(u,v) \in \Psi$ is subsequently computed as $\kappa(u,v) \cdot \min(1, w(v)/w(u))$, where the function~$w\colon \Omega \to \mathbb{R}^+$ can optionally be overridden to enable the Metropolis rule (see Algorithm~\ref{alg:randomwalk}).
The construction of the state graph relies on connectedness of the state graph, which is given whenever the Markov chain is irreducible. The construction of the state graph is a sequential process, consuming $O(|\Omega| + |\Psi|)$ time and memory.

\paragraph{Computation of the Total Mixing Time}

We compute the total mixing time $\tau(\epsilon)$ of a state graph based on the following idea: Let~$P$ be the transition matrix of an ergodic Markov chain and let $a \in \Omega$ be an initial state. One step of the random walk changes the probability distribution $p^{(t)}_a$ to $p^{(t+1)}_a = p^{(t)}_a \cdot P$. Iterating this argument, the distribution after $t$ steps is given by
\begin{equation}
p^{(t)}_a = p_{a}^{(0)} \cdot P^t, \label{eqn:steprandomwalk}
\end{equation}
where $p_{a}^{(0)}$ is a row vector whose components are $p_{a}^{(0)}(a) = 1$ for the initial state $a$ and $p_{a}^{(0)}(i) = 0$, for $i \not= a$. So, each row of the matrix $P^t$ gives the distribution to a certain $p_a^{(t)}$ for $a\in \Omega$. The total mixing time $\tau(\epsilon)$ can therefore be computed from~$P$ by iterated matrix multiplication.
The principle can be described as follows: We transform the adjacency list representation of $\Gamma$ into its adjacency matrix representation. The entries $P_{i,j}$ of this matrix represent the transition probabilities for going from state $i$ to state $j$, so this matrix is the transition matrix described above. 
Since the total variation distance decreases monotonically with $t$, we can find $\tau(\epsilon)$ with a two-step procedure. 
\begin{enumerate}
\item Starting with $i=0$, we iteratively increment $i$ until $\max_{a \in \Omega} || p_a^{(2^i)} - \pi || \leq \epsilon$. This requires, in fact, just a sequence of $i = \lceil \log_2(\tau(\epsilon)) \rceil$ matrix squaring operations to compute the matrices $P^2, P^4, P^8, \ldots, P^{2^i}$. At the end of step one, we know that the total mixing time $\tau(\epsilon)$ lies in the range $2^{i-1} < \tau(\epsilon) \leq 2^i$. 
\item We use binary search to find $\tau(\epsilon)$. We define two variables $l = 2^{i-1}$ and $r = 2^i$ representing respectively lower and upper bounds on $\tau(\epsilon)$. In a sequence of $\log_2(r-l) = \lceil \log_2(\tau(\epsilon)) \rceil - 1$ iterations, we half the search space in each step by computing the total variation distance $\max_{a \in \Omega}||p_a^{(m)} - \pi||$ with $m=\lfloor(l+r)/2\rfloor$ from $P^m$. Although $P^m$ could, in principle, be computed from $P$ with binary exponentiation, we save additional matrix multiplications by reusing $P^l$ to compute $P^m = P^l P^{m-l}$. Maintaining the invariant $\max_{a \in \Omega}||p_a^{(l)} - \pi|| > \epsilon \geq \max_{a \in \Omega} ||p_a^{(r)} - \pi||$, we update the lower bound $l$ or the upper bound $r$, depending, whether the total variation distance is larger than $\epsilon$ or not. We stop when $l \geq r$. The value of $r$ is the total mixing time.
\end{enumerate}
Computing the total mixing time requires $O(|\Omega|^2)$ memory 
and $O(\log(\tau(\epsilon)) \cdot |\Omega|^{\omega})$ time, where $\omega$ is the exponent of the matrix multiplication algorithm. (In most libraries, $\omega$ is equal to $3$. However, the matrix multiplication implementation can be exchanged by using another implementation when building the library.)

\begin{figure}[h]
	\includegraphics[width=\textwidth]{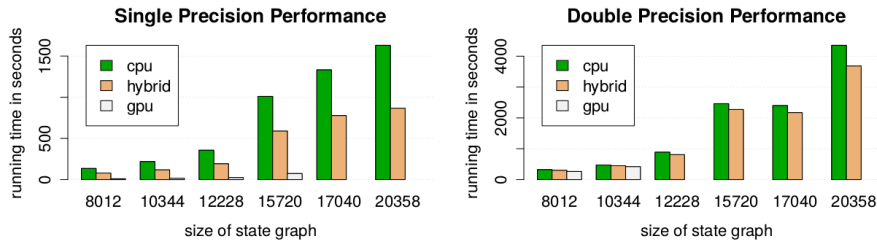}
   \caption{\textbf{Single and double precision performance of the total mixing time computation.} The charts show the running time for the computation of the total mixing time on the example of five state graphs of size $8012$ to $20358$. Due to the relatively small amount of GPU memory on our test system, only the first four (respectively two) state graphs could be processed by the GPU implementation in single precision mode (respectively double precision mode). The running times were measured on an Ubuntu 14.04 system with a Intel Xeon E3-1231, NVIDIA GeForce GTX 970 (4 GB GPU memory) and 16 GB of main memory, using \textsl{gcc} in version 4.8.4 and \textsl{CUDA} in version 7.0.  } 
   \label{Fig:performance}
\end{figure}

Computing the total mixing time is currently by far the most time and memory consuming operation in our library. Due to quadratic memory requirement, it is applicable only for small state graphs with $|\Omega| \leq 30000$ (depending on main memory). Since this method is also very time-consuming, we provide three implementations, which can be used if the appropriate hardware is available:
\begin{enumerate}
\item[a)] A classical CPU implementation on the basis of \textsl{openBLAS} \cite{openblas}, which runs in parallel on multi-core CPUs. This method should be used for very small instances, or if no \textsl{CUDA}-capable GPU is available.
\item[b)] A \textsl{cuBLAS} \cite{cublas} based GPU implementation which can be significantly faster than the CPU implementation. Because of the lack of memory on most GPUs, this implementation is designed for small state graphs with $|\Omega| \leq 15000$ (depending on GPU memory).
\item[c)] A CPU-GPU hybrid implementation using the \textsl{cuBLASXt} \cite{cublasxt} library for matrix multiplication. This implementation should be the method of choice if a \textsl{CUDA} capable GPU is available. This method is typically faster than the CPU implementation and has the additional advantage that the CPU is free to be used for other computations.
\end{enumerate}
The three implementations share the same asymptotical running time, but differ greatly in practice (see Fig.~\ref{Fig:performance} for a comparison of performance). Since most of the work is matrix multiplication, this comparison almost directly maps to a performance comparison between the corresponding libraries for matrix multiplication. The $O(\log(\tau(\epsilon)))$ invocations of the total variation distance computation, each with time complexity of $O(|\Omega|^2)$, contribute only a little to the running time.

\paragraph{Computation of the Spectral Bound}

The difference between the largest and the second largest eigenvalue $(1-\lambda_{\max})$ of a Markov chain's transition matrix is known as the \emph{spectral gap} and is of central interest in mixing time analysis. To compute this quantity, we use the well-known \textsl{ARPACK++} \cite{arpack} library to compute the two real eigenvalues with the largest magnitude of the transition matrix $P$. In case of non-symmetric transition rules, we first transform the transition matrix $P$ into the symmetrical matrix $P_{sym}$ via Equation~(\ref{eqn:symmetrical}). After computing both eigenvalues with the largest magnitude, we use the second eigenvalue in order to compute the upper and lower spectral bound via Inequalities~(\ref{eqn:upper_spectral_bound}) and~(\ref{eqn:lower_spectral_bound}).  Since this method requires only a sparse representation of the transition matrix, it is also applicable for large state graphs.

\paragraph{Computing the Congestion Bound}

As stated in the introduction, we are also interested in the quality of other bounding techniques. In particular, we are interested in the canonical path method which is often used to gain theoretical bounds. We implemented a method that can be used to gain an upper bound on the mixing time by computing the maximal congestion of a user-definable path construction scheme. Four each pair of states $(u,v) \in \Omega \times \Omega$ we apply the path construction scheme to construct a path $p$ from state $u$ to state $v$. The congestion of each transition arc lying on this path is increased by $|p| \pi(u) \pi(v)$ (see Equation~(\ref{eqn:maximumLoading})). Since the construction of $|\Omega|^2$ paths can be done completely independently of each other, we use \textsl{OpenMP} to construct all paths in parallel. The maximal congestion of any transition arc is used for computing the upper bound via Inequality~(\ref{eqn:flow2}). This method has a time complexity of $O(|\Omega|^2)$ and a memory complexity of $O(|\Psi|)$.

\paragraph{Network Analysis}

As state graphs can be seen as weighted directed graphs, all kinds of graph algorithms can be integrated into \textsl{marathon}. As an example, we added functions for computing the diameter of a state graph, as well as functions for computing the average path length. 

\subsection*{Exemplary Markov-Chains}

To demonstrate possible applications of \textsl{marathon}, we will describe a set of experiments in the following section.
In our experiments, we focus on two famous sampling problems. The problem of uniformly sampling a \emph{perfect matching} in a \emph{bipartite graph} and of uniformly sampling a \emph{bipartite graph realization} for given \emph{vertex degrees}. Both problems are based on bipartite graphs. 
A \emph{bipartite graph} $G=(U,V,E)$ is an undirected graph with disjoint vertex sets $V$ and $U$ and a set of edges $E \subseteq U \times V$, connecting vertices from $U$ with $V$. We will refer to the vertices of $U$ as $u_i$, where $i$ ranges from $1$ to $n := |U|$ and to the vertices of $V$ as $v_j$, where $j$ ranges from $1$ to $n' := |V|$. We analysed four different chains that can be used for these sampling problems. To make this article self-contained, we give a introduction to the chains.

\subsection*{Markov Chains for Sampling of Perfect Matchings}
\label{sec:matching}


A \emph{perfect matching} in a bipartite graph $G=(U,V,E)$, with $n=n'$, is a subset $M \subseteq E$ of edges such that a) all edges in $M$ are non-adjacent, and b) $|M|=n$. 
Uniformly sampling a perfect matching is the problem of choosing one perfect matching from the set of all perfect matchings of $G$ uniformly at random.
We describe two important Markov chains which are known to solve the problem of uniformly sampling perfect matchings.
These chains origin from applications in statistical physics and are widely known in the field of Markov chain analysis, so they make good examples for our analysis.
Both Markov chains use the set of near-perfect matchings as auxiliary states. A \emph{near-perfect matching} is a subset of non-adjacent edges from $E$, but with one edge less than a perfect matching. So, being specific, the Markov chains used for sampling are based on the set $\Omega_M$ of perfect and near perfect matchings in $G$. 

\paragraph{Matching Chain One}

Broder \cite{broder86} introduced a Markov chain which can be described as follows. In state $M \in \Omega_M$ we pick a candidate state $M'$ (line three in Algorithm~\ref{alg:randomwalk}) by choosing an edge $e=\{u,v\}$ from $E$ uniformly at random. One of five cases occurs:
\begin{enumerate}
\item If $M$ is a perfect matching and $e \in M$, then remove $e$ from $M$ and select $M' \gets M \setminus \{e \}$.
\item If $M$ is a near-perfect matching and $e \not\in M$, then add $e$ to $M$ and select $M' \gets M \cup \{ e \}$.
\item If $M$ is a near-perfect matching, $u$ is not matched in $M$, and an edge $e'=\{ w,v \} \in M$ exists, then remove $e'$ from $M$, add $e$ and select $M' \gets \left( M \setminus \{ e' \} \right) \cup \{ e \}$.
\item Symmetrically, if $M$ is a near-perfect matching, $v$ is not matched in $M$, and an edge $e'=\{ u,z \} \in M$ exists, then remove $e'$ from $M$, add $e$ and select $M' \gets \left( M \setminus \{ e' \} \right) \cup \{ e \}$.
\item In all other cases, stay at $M$.
\end{enumerate}
The state $M'$ is proposed as a candidate state with proposal probability $\kappa(M,M') = 1/|E|$ in line three of Algorithm~\ref{alg:randomwalk}. The proposed state $M'$ is accepted or refused by line four of Algorithm~\ref{alg:randomwalk}, using unit weights. So, this chain converges to a uniform stationary distribution. In the remainder of this article, we refer to this Markov chain as \emph{Matching Chain One}. Jerrum and Sinclair \cite{Jerrum:1989:AP:76071.76077} proved that Matching Chain One is rapidly mixing for a graph class where the ratio of near-perfect matchings $\mathcal{N}(G)$ and perfect matchings $\mathcal{M}(G)$ in $G$ is polynomially bounded. Being specific, its total mixing time can be bounded from above:
\begin{equation}
\tau(\epsilon) \leq 16^2|E|^2\left(\frac{|\mathcal{N}(G)|}{|\mathcal{M}(G)|}\right)^4\ln(|\Omega_M|\cdot\epsilon^{-1}).
\end{equation}
The class of \emph{dense} bipartite graphs, i.e. bipartite graphs $G=(U,W,E)$ with $|U|=|W|=n$ and with minimal vertex degree of at least $n/2$, was shown as a class for which the ratio of near-perfect and perfect matchings is polynomially bounded \cite{Jerrum:1989:AP:76071.76077}. For this class of graphs, the following upper bound on the total mixing time can be gained \cite{Diaconis99statisticalproblems}:
\begin{equation}
\tau(\epsilon) \in O(n^7\ln (|\Omega_M|\cdot \epsilon^{-1})). \label{eqn:matching1dense}
\end{equation}
A main problem of Matching Chain One is the fact that it does not only generate perfect matchings but also \emph{near-perfect} matchings. In many graphs with $2n$ vertices the ratio of the number of near-perfect and perfect matchings cannot be bounded by a polynomial which is dependent on $n$. In such cases, the sampling of a perfect matching needs an exponential number of trials to see a perfect matching at all.

\paragraph{Matching Chain Two}

One way to overcome the problem of an exponential total mixing time has been given by Jerrum, Sinclair and Vigoda \cite{JerrumSinclairVigoda04}. Their main idea is to use a carefully chosen weight function $w$ to sample from a non-uniform distribution. This way, they gained a rapidly mixing Markov chain which can be used to sample perfect matchings from any given bipartite graph. In addition, the transition rules were changed to the following: In a state $M\in \Omega_M$:

\begin{enumerate}
\item If $M$ is a perfect matching, we choose an edge $e=\{u,v\}$ from $M$ uniformly at random. Remove $e$ from $M$ and select $M' \gets M \setminus \{e \}$.
\item If $M$ is a near-perfect matching where $u$ and $v$ are unmatched vertices, we choose a vertex $z \in U \cup V$ uniformly at random.
	\begin{enumerate}
		\item[2a)] If $z$ is one of the unmatched vertices $u$ and $v$ and $e=\{ u,v \} \in E$, then add $e$ to $M$ and select $M' \gets M \cup \{ e \}$.
		\item[2b)] If $z \in V$, $\{ u,z \} \in E$ and $\{ x,z \} \in M$, then remove the edge $\{ x,z \}$ from $M$, add $\{ u,z \}$ and select $M' \gets \left( M \setminus \left\{ \{ x,z \} \right\} \right) \cup \left\{ \{ u,z \} \right\}$.
		\item[2c)] If $z \in U$, $\{ z,v \} \in E$ and $\{ z,y \} \in M$, then remove the edge $\{ z,y \}$ from $M$, add $\{ z,v \}$ and select $M' \gets \left( M \setminus \left\{ \{ z,y \} \right\} \right) \cup \left\{ \{ z,v \} \right\}$.
	\end{enumerate}
\end{enumerate}
The proposal probability $\kappa(M,M')$ for two states $M, M' \in \Omega_M$ is therefore $1/n$ when moving between perfect and near-perfect matchings and $1/(2n)$ when moving between near-perfect matchings. 
In connection with Algorithm~\ref{alg:randomwalk}, this Markov chain converges to a stationary distribution $\pi$ which is proportional to the weight function $w$. Defining $\mathcal{M}$ as the set of perfect matchings of $G$, and $\mathcal{N}_{u,v}$ as the set of near perfect matchings in $G$ where $u$ and $v$ are unmatched, the weight function suggested in \cite{JerrumSinclairVigoda04} is defined as:
\begin{equation}
w(M) = \begin{cases}
1,& M \in \mathcal{M}\\
|\mathcal{M}| / |\mathcal{N}_{u,v}|, & M\in \mathcal{N}_{u,v}.
\end{cases}
\end{equation}
Knowing the values of $|\mathcal{M}|$ and $|\mathcal{N}_{u,v}|$, the total mixing time of this chain can be bounded from above by the following polynomial \cite{Bezakova06acceleratingsimulated}:
\begin{equation}
\tau(\epsilon) \in O(n^4 \ln((\pi_{min}\cdot\epsilon)^{-1}).
\end{equation}
Unfortunately, $|\mathcal{M}|$ and $|\mathcal{N}_{u,v}|$ are usually not known in practice. Jerrum, Sinclair and Vigoda  gave a description of a procedure for approximating these quantities in polynomial time \cite{JerrumSinclairVigoda04}. In our experiments though, we knew the exact values since we had a list of all states. In combination with Algorithm~\ref{alg:randomwalk}, this Markov chain is rapidly mixing for all bipartite graphs and can be used for sampling of perfect matchings. We refer to this chain as \emph{Matching Chain Two}.
	
\subsection*{Markov Chains for Sampling of Bipartite Graph Realizations}

The \emph{bipartite graph realization problem} is to find, for a pair of non-increasing integer sequences $(a_1,\dots,a_{n})$, $(b_1,\dots,b_{n'})$ with $a_i,b_i >0$, a bipartite graph $G=(U,V,E)$, without loops and parallel edges, such that the vertex degrees of $u_i \in U$ matches the $a_i$ and that the vertex degrees of $v_i \in V$ match the $b_i$. Such a graph $G$ is then called \emph{bipartite graph realization} or, shorter still, \emph{realization}. The set of all realizations of a given degree sequence pair makes the sampling set $\Omega_R$ for this sampling problem.  The bipartite graph realization problem is so important in many fields that it was reinvented several times. It is also known as the problem of \emph{contingency tables with given marginal sums} or, as \emph{matrices with fixed row and column sums}.
As a direct application of random bipartite graphs, scientists often use a sampled realization as a null model to prove statistical hypotheses. Since sampling of random bipartite graphs is important in a wide range of fields, we chose it as a second example for our experiments.

In contemporary literature, one can find two possible Markov chains that solve the problem of uniform sampling of a bipartite graph realization. One of these approaches is simply to transform it into the perfect matching sampling problem~\cite{Tutte52}. This way, it is known that the running time of the sampling procedure is bounded polynomially. However, the very simple and intuitive \emph{switch chain}, presented by Kannan et al. \cite{Kannan97}, is widely used in practice.

\paragraph{Switch Chain One}

At a state $G = (U,V,E) \in \Omega_R$, choose four integers $i,j,k,l$ with $1 \leq i \leq k \leq n$ and $1 \leq j \leq l \leq n'$ uniformly at random. One of three cases occur: 
\begin{enumerate}
\item If the edges $e_1 = \{ u_i, v_j \}$ and $e_2 = \{ u_k, v_l \}$ exist in $E$ but the edges $e_1' = \{ u_i, v_l \}$ and $e_2' = \{ u_k, v_j \}$ do not, then select $G'=(V,U,E')$ with $E' \gets \left( E \setminus \left\{ e_1, e_2 \right\} \right) \cup \left\{ e_1', e_2' \right\}$.
\item Symmetrically, if $e_1 = \{ u_i, v_j \}$ and $e_2 = \{ u_k, v_l \}$ both do not exist in $E$ but the edges $e_1' = \{ u_i, v_l \}$ and $e_2' = \{ u_k, v_j \}$ exist, select $G'=(V,U,E')$ with  $E' \gets \left( E \setminus \left\{ e_1', e_2' \right\} \right) \cup \left\{ e_1, e_2 \right\}$.
\item In all other cases, select $G' \gets G$.
\end{enumerate} 
Such a \emph{switch}, i.e. changing the endpoints of two edges, does not change the degree sequence and thus constructs a new bipartite graph realization of $S$. It turns out, all graph realizations can be generated by starting with a given graph realization and applying a number of switches \cite{Petersen1891}.
The proposal probability $\kappa(G, G')$ of a non-loop transition is $\frac{2}{n(n+1)}\cdot\frac{2}{n'(n'+1)}$.
Using unit weights $w(x)=1$ for all $x \in \Omega_R$ Algorithm~\ref{alg:randomwalk} converges to a uniform stationary distribution. Kannan et al. \cite{Kannan97} proved polynomially bounded mixing times of this chain for regular sequence pairs. Mikl{\'o}s et al. \cite{Miklos13} extended the proof to half-regular sequence pairs. For all other sequence classes, the problem of the rapid mixing property is still open. We will refer to this Markov chain as \emph{Switch Chain One}. 
Recently, Greenhill gave a proof that the total mixing time of the directed, non-bipartite version of Switch Chain One is bounded by the following polynomial when the largest degree $d_{\max}$ of the input sequence lies in the range $3 \leq d_{\max} \leq \frac{1}{4} \sqrt{|E|}$, where $|E|$ is the number of edges in the realization \cite{DBLP:conf/soda/Greenhill15}:
\begin{equation}
\tau(\epsilon) \leq \frac{1}{10} d^{14}_{\max} |E|^9 ( |E| \ln |E| + \ln(\epsilon^{-1})).
\end{equation}

\paragraph{Switch Chain Two}

A modification of Switch Chain One, given by Berger et al.~\cite{Berger:2010} for directed graph sequences, can also be used for the bipartite version. The main idea is to reduce the number of transition loops in the state graph. To avoid non-convergence, an additional additional artificial edge $\{ u_0, v_0 \}$, which connects two additional vertices $u_0$ and $v_0$, is added to the set of edges.
In a state $G=(U,V,E) \in \Omega_R$ select a pair of non-adjacent edges $e_1 = \{ u_i, v_j \}$ and $e_2 = \{ u_k, v_l \}$ from the set $E \cup \left\{ \{ u_0, v_0 \}  \right\}$.
\begin{enumerate}
\item If, $e_1 = \{ u_0, v_0 \}$ or $e_2 = \{ u_0, v_0 \}$, select $G' \gets G$.
\item If $e_1' = \{ u_i, v_l \} \not\in E$ and $e_2' = \{ u_k, v_j \}  \not\in E$ , then switch the edges $(e_1, e'_1, e_2, e'_2)$ and select $G'=(V,U,E')$ with $E' \gets \left( E \setminus \left\{ e_1, e_2 \right\} \right) \cup \left\{ e_1', e_2' \right\}$.
\item In all other cases, select $G' \gets G$.
\end{enumerate}
Without the additional edge, a problem would occur if no adjacent edge pairs existed for each graph realization of a sequence; then, the state graph would not possess loops. 
Similar arguments like those of Ref.~\cite{Berger:2010} show that \rem{the}\add{this} chain converges to the uniform distribution.
This chain reduces the average probability of transition loops and, as it will turn out, the average total mixing time. However, the asymptotical total mixing time is the same as for Switch Chain One. We will refer to this chain as \emph{Switch Chain Two}.

\section*{Experiments}
\label{Section:experiments}

We demonstrate our library to experimentally analyse the mixing time behaviour of the Markov chains. The goal of our experiments is to quantify the gap between the total mixing time and the various bounding methods. We are asking the question, whether the bounding techniques are tight or if there is a huge gap between total mixing time and its bounds. The latter case would support the thesis that the total mixing time actually is  far smaller than known in theory. This would be a welcomed message for any programmer who wants to implement a sampling method and does not want to choose a highly polynomial (or even exponential) number of steps. We want to make some general remarks before beginning to describe our experiments.

\paragraph{The choice of $\pmb{\epsilon}$}

Recall, that the total mixing time $\tau(\epsilon)$, as well as its bounds depends on the value of $\epsilon$ which defines the  distance to the stationary distribution at the end of the random walk. For the purpose of our experiments, we choose $\epsilon=10^{-3}$. Furthermore, for other values of $\epsilon$ we get similar results. 

\paragraph{Applying canonical paths}

To compute the congestion bounds, we use the path construction schemes from the original proofs  and apply them to concrete state graphs. As those path construction schemes were used to gain theoretical bounds without full knowledge of a Markov chain's structure, applying them to concrete state graphs shows how sharp such bounds could be if full structural information were available. Consequently, the  bounds in general cannot be better than the upper bounds that are gained when applying the scheme on an actual state graph. To understand, how sharp the theoretical bounds can be, we analyse the differences to the actual total mixing times. In the case of matching chains, we use the construction scheme presented in \cite{Jerrum:1989:AP:76071.76077} which was used to gain the upper bound given in Equation~\ref{eqn:matching1dense}. In case of the switch chains we apply the definition of the canonical paths from \cite{Kannan97}. For the definition of the path construction schemes, we refer to the original articles. We tried as close as possible to implement the original canonical path construction rules. However, the definition often relies on an arbitrary order of the vertices, or orderings of cycles. It turned out that changing these orders has an effect on the quality of the congestion bound. However, as the proofs for the upper bounds are valid for all orderings, this is not a problem for our observations.

\begin{figure}[h!]
\begin{center}
\includegraphics[width=\textwidth]{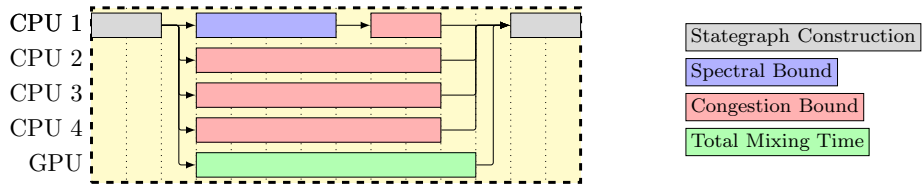}
\vspace{1mm}
\caption{\textbf{Typical hardware usage for our experiments.} At first, the state graph is being constructed as a sequential task. Afterwards, the properties of the state graph are being computed in parallel. When each property is computed the next instance is processed in the same manner.}
\label{Fig:hardware_occupancy}
\end{center}
\end{figure}

\paragraph{Hardware Usage}
Since we wanted to compute the properties of a large amount of instances, we decided to compute as many properties of a state graph as possible in parallel. As shown before, the computation of the total mixing time is the bottleneck for our computations. For this reason, we decided to transfer this task to the GPU whenever possible. This has the advantage, that the free CPU cores can be used for the computation of other quantities, while the GPU is occupied. With this strategy, we achieve a large throughput of instances. Fig.~\ref{Fig:hardware_occupancy} shows the hardware usage while computing the properties of a Markov chain instance.

\subsection*{Enumeration Experiment}

In our first experiment, we wanted to gain an impression concerning the dimension of the total mixing time and its bounds. 

\paragraph{Experimental Setup}
Firstly,  we enumerated all small input instances up to a given limit for each chain. In particular, we enumerated 19,378 bipartite degree sequence pairs with up to 6+6 vertices as input for the switch chains and 89,242 connected non-isomorphic bipartite graphs with 6+6 vertices as input for the matching chains. While the first task was done with a simple backtracking procedure, the latter task was done by the command line tool \textsl{nauty} \cite{McKay201494}. These combinatorial objects served as input instances for the Markov chains.  Even though the size of input is small, the resulting state graphs can be large. For example, the largest state graph processed in this experiment had 297,200 states and corresponds to the regular degree sequence pair $(3,3,3,3,3,3)$,$(3,3,3,3,3,3)$. Since both the number of instances, as well as the size of the state graphs grow exponentially, we were not able to perform such an experiment for a much larger instance size.
For each instance, we constructed the corresponding state graph. For all state graphs with a maximum of 20,000 states, we computed the following properties: a) total mixing time, b) spectral bound, and c) canonical path congestion bound. Although we could easily compute the latter bounds for much larger state graphs, we did not do so, since we need the total mixing time for comparison.

\begin{figure}[h!]
\begin{center}
\includegraphics[width=\textwidth]{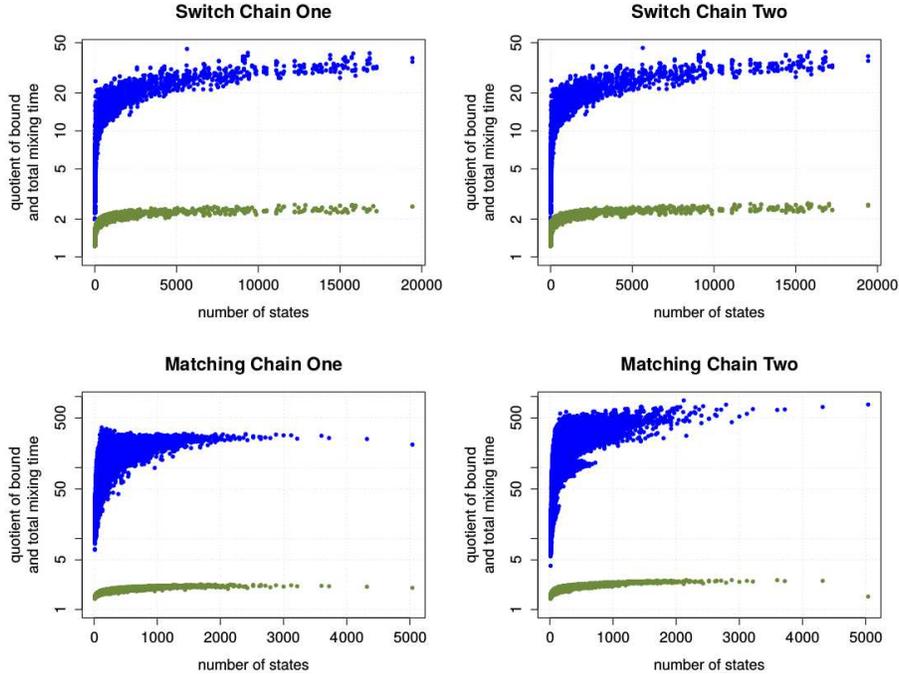}
\caption{\textbf{The quality of the upper bounds.} The quotient of congestion bound (blue) and total mixing time, respectively upper spectral bound (green) and total mixing time versus the size of the corresponding state graph. 
}
\label{Fig:size_vs_bounds}
\end{center}
\end{figure}

\begin{figure}[h!]
\begin{center}
\includegraphics[width=\textwidth]{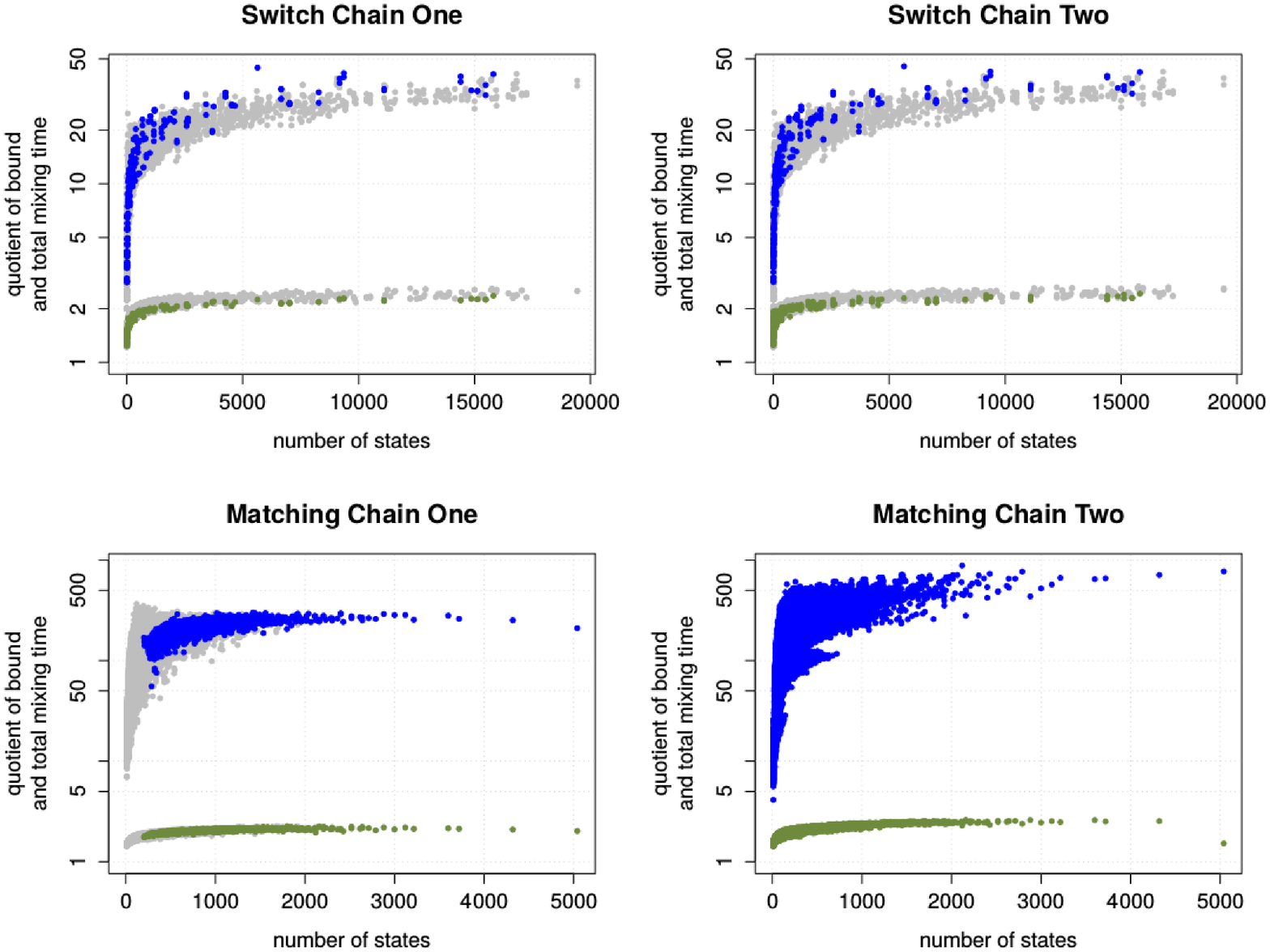}
\caption{\textbf{The quality of the upper bounds for rapidly mixing instances.} The results of Fig.~\ref{Fig:size_vs_bounds} filtered to highlight instances with known polynomial mixing time. Instances with no known polynomial bound are coloured gray.}
\label{Fig:size_vs_bounds_rapidly_mixing}
\end{center}
\end{figure}

\paragraph{Results} In Fig.~\ref{Fig:size_vs_bounds}, we show the results of the enumeration experiment. Each input instance corresponds to two data points, showing the ratio of the upper spectral bound (green), and congestion bound (blue) with the total mixing time. We immediately observe that the spectral bound is very close to the total mixing time for all instances, in contrast to the congestion bound. More precisely, in case of Matching Chain Two, the congestion bound is up to 800 times larger than the total mixing time. Similar observations can be made for the other chains.
In Fig.~\ref{Fig:size_vs_bounds_rapidly_mixing}, we highlight the instances which are known to have a polynomial congestion bound. In the case of the switch chains, these are regular and half-regular degree sequences. In the case of Matching Chain One, these are dense bipartite graphs. Matching Chain Two is known to be rapidly mixing for all instances. We find that the observations can also be confirmed within this set. 



\subsection*{Scaling Experiment}

We wanted to know whether the observed effects remain valid for larger instances. The main problem is, that the number of instances for our Markov chains, both degree sequences and bipartite graphs, grows exponentially with input size. So, a complete enumeration becomes infeasible when the instance size grows. To gain some insights, we picked selected instances for the switch chains and scaled them to larger size.

\paragraph{Experimental Setup}

We picked a half-regular sequence pattern $(n-1,n-2,2,1)$, $(2,2,\ldots , 2)$, where the number of two's in the second sequence is $n$. The parameter $n$ can be used to scale the instance to different sizes. In doing so, a bipartite graph which realizes this degree sequences has $n+4$ vertices and $2n$ edges. These sequence pairs are half-regular and so, it is known that the switch chains are rapidly mixing. For each $n$ between 4 and 50, we constructed the corresponding state graph and computed the total mixing time and its bounds. Again, due to the large memory consumption, we stopped computing the total mixing time when $|\Omega|$ exceeded 20,000. To predict missing values of total mixing time for larger state graphs up to about 600,000 states, we used a linear model based on the following observation.

\begin{figure}[h!]
\begin{center}
\includegraphics[width=\textwidth]{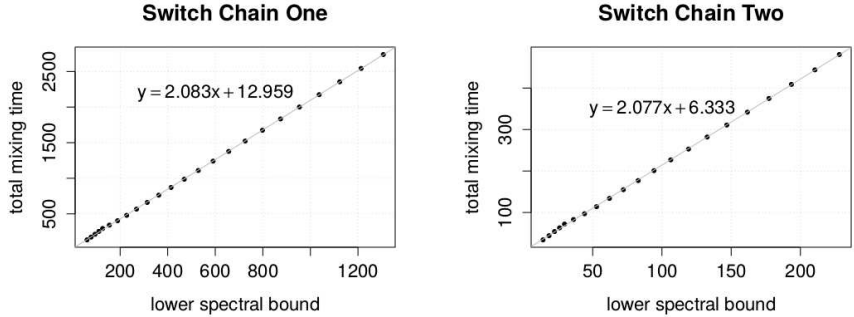}
\caption{ \textbf{Relationship between the lower spectral bound and the total mixing time.} The total mixing time is shown in connection to a corresponding lower spectral bound for sequence pairs of the form $(n-1,n-2,2,1)$, $(2,2,\ldots , 2)$. We use the displayed formulas to predict missing values for total mixing time.}
\label{Fig:half_regular2_aux}
\end{center}
\end{figure}

\begin{figure}[h!]
\begin{center}
\includegraphics[width=\textwidth]{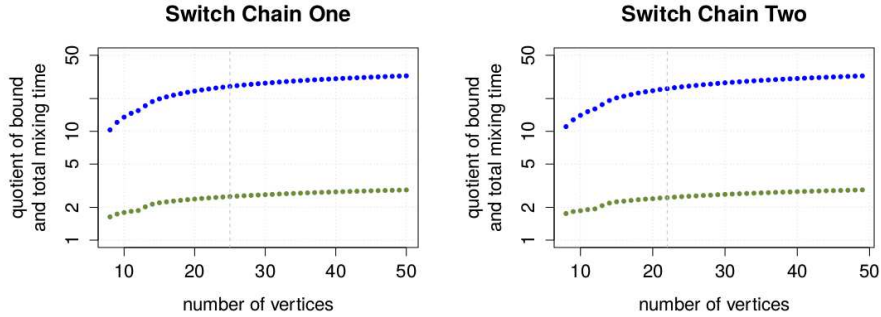}
\caption{\textbf{Quality of the upper bounds with growing instance size.} The quotient of congestion bound (blue), respectively upper spectral bound (green) with total mixing time is shown as a function of the instance size. A dotted line shows the point where we stop computing the total mixing time and start using the lower spectral bound approximation.}
\label{Fig:half_regular2_quotient}
\end{center}
\end{figure}

\begin{figure}[h!]
\begin{center}
\includegraphics[width=\textwidth]{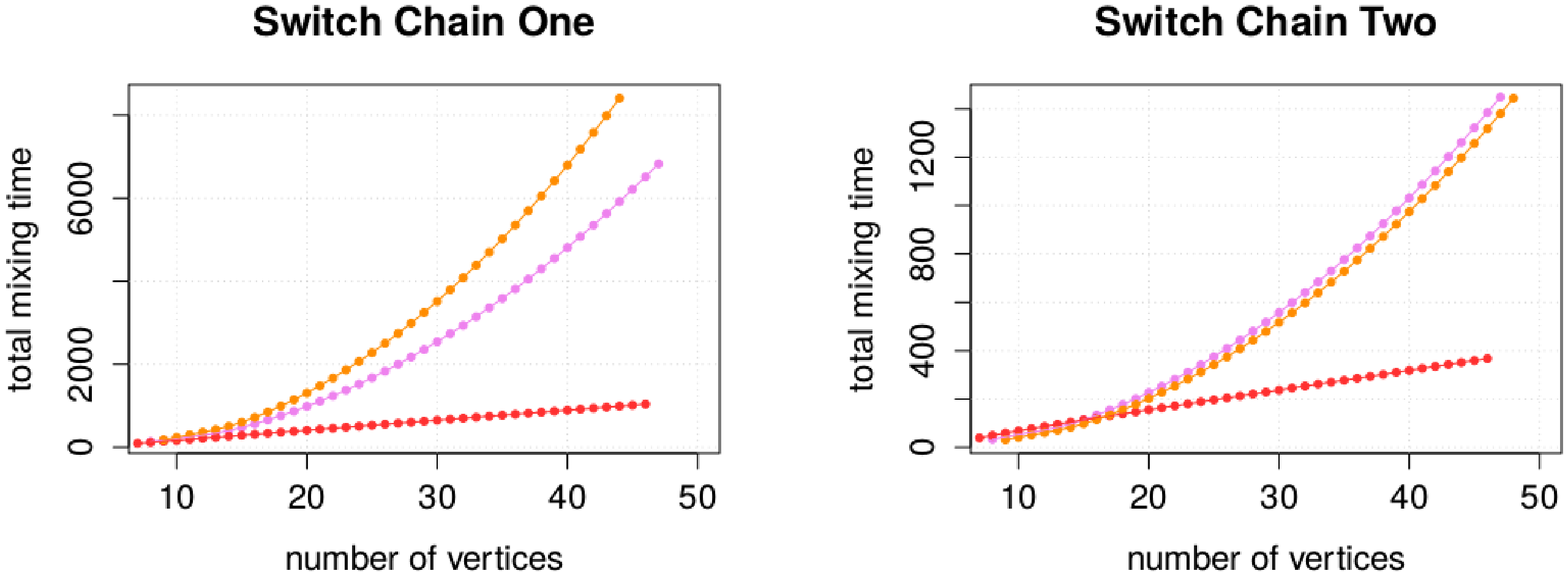}
\caption{\textbf{Growing asymptotic for the total mixing time of half-regular sequence pairs.} The observed mixing time curves for the degree sequence pairs of type A are coloured violet, of type B red, and of type C orange.}
\label{Fig:half_regular_instances}
\end{center}
\end{figure}

\paragraph{Results}
We observed that the lower spectral bound seems to have a linear relationship with the total mixing time. This allowed us to predict the missing total mixing time values using a simple linear model (see Fig.~\ref{Fig:half_regular2_aux} for details).
To quantify the gap between the bounds and total mixing time, we again divided each bound through the corresponding total mixing time or its predicted value. Fig.~\ref{Fig:half_regular2_quotient} shows the result of this experiment. We observed that the gap between the congestion bound and the total mixing time grows with input size~$n$. Furthermore, we observed that the same is true for the upper spectral bound, but in a much slower way. 
We repeated this experiment with other half-regular sequence patterns and observed similar effects. In particular, we changed our sequence pattern $(n-1,n-2,2,1)$, $(2,2,\ldots, 2)$, which we call type A, slightly to the sequences patterns $(n-1,n-2,3)$, $(2,2,\ldots, 2)$ (type B) and $(n-1,n-2,1,1,1)$, $(2,2,\ldots, 2)$ (type C). Fig~\ref{Fig:half_regular_instances} shows the, partially estimated, total mixing time for each sequence type. We observed that each type has its own growing asymptotic. The data suggests, in the case of Switch Chain One, the total mixing time of sequence type A grows with $O(n^{2.27})$, of type B with $O(n^{1.16})$, and of type C with $O(n^{2.43})$. In the case of Switch Chain Two, we observe similar effects.
These results are interesting because these bounds are lower bounds on the mixing time of the switch chains for all half-regular sequence pairs. Since two of them are clearly super-linear, a random walk with a linear number of steps must ultimately fail and will result in a non-random sample.

\subsection*{Loop Reduction Experiment}

\begin{figure}[h!]
\begin{center}
\includegraphics[width=\textwidth]{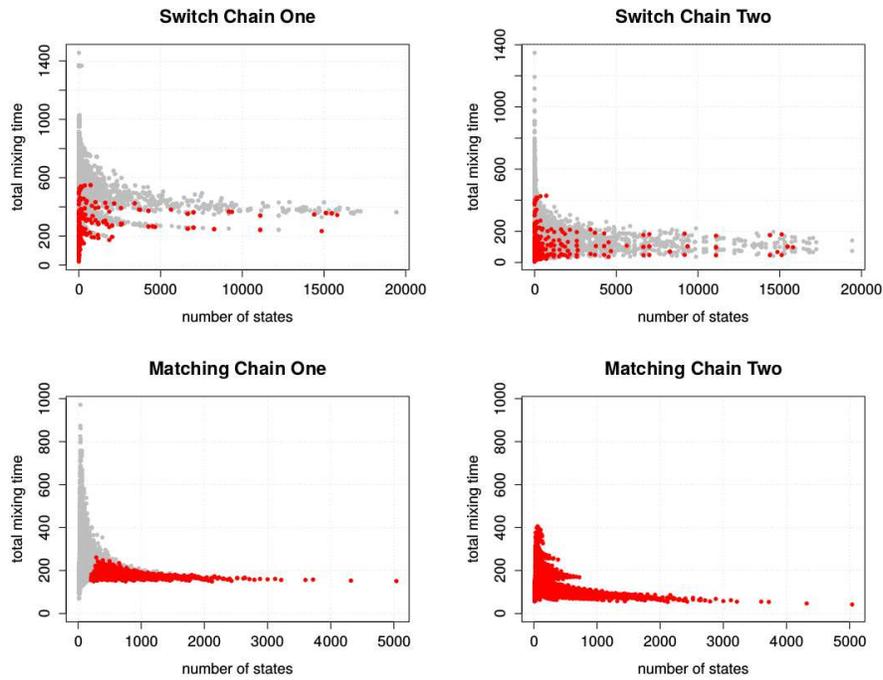}
\caption{\textbf{Number of states versus total mixing time.} Each data point represents one input instance from our set of input instance. Instances with known polynomial mixing time are highlighted red.}
\label{Fig:mix_vs_size}
\end{center}
\end{figure}

From the data of the enumeration experiment, we observed that the total mixing time does not correlate with the size of the state graph (see Fig.~\ref{Fig:mix_vs_size}). Instead, the largest mixing time appears at input instances corresponding to small state graphs. 
For example, the sequence pair $(6,6,6,6,5,5)$,$(6,6,6,6,5,5)$, possesses just two different states but is beneath the instances with largest total mixing time in both switch chains. To explain this effect, consider sequence pairs of the general form $(n,n,\ldots, n-1, n-1)$, $(n,n,\ldots, n-1, n-1)$, where $2n$ is the number of vertices in the corresponding bipartite graph. It is apparent that such sequence pairs posses two different realizations, meaning, their state graph have two states. Now consider Switch Chain One; for a given $n$, the number of ways to choose $i,j,k,l$ is $(n\cdot(n+1)/2)^2 \in O(n^4)$. Due to the large number of edges in the sequence pairs, all but one of these choices result in loops. As a consequence, the expected number of steps, just to get from one state to the other, is $O(n^4)$, which is immediately a lower bound for the mixing time of Switch Chain One. The fact that an instance with polynomially bounded total mixing time makes the largest mixing time over all instances is likely an effect of the small input size. 
To better understand this effect, we decided to further reduce the influence of the loops. 

\paragraph{Experimental Setup}

We repeated the enumeration experiment but made some artificial modifications on each state graph before computing its properties. For each state graph we determined the minimal loop probability over all states $P_{\min} := \min_{i \in \Omega} P_{ii}$. We wanted to reduce the transition probability of each loop in the state graph by this quantity. However, to avoid the removal of all loops and possible problems with convergence, we removed only 99\% of this probability from the loops. This way, we gained new transition probabilities of 
\begin{equation}
P'_{ii} \gets P_{ii} - .99P_{\min} \textnormal{ for all } i \in \Omega.
\end{equation}  
The amount of $.99P_{\min}$ which was reduced from the probability mass of each state, was restored to the network through rescaling of the remaining transition arcs:
\begin{equation}
P'_{ij} \gets \frac{P_{ij}}{1 - .99  P_{\min}} \textnormal{ for all } i,j \in \Omega.
\end{equation}
By subtracting the same amount of probability from each loop, we kept the symmetry of the transition matrix so that the chain's stationary distribution was still uniform. The result was a state graph with the same structure but drastically reduced loop probability. In such state graphs, the mixing times are much more dependent on the structure of the state graph than on the number of the loops. We repeated the computation of the total mixing time and its upper bounds with those modified state graphs for Switch Chain One.

\paragraph{Results}

\begin{figure}[h!]
\begin{center}
\includegraphics[width=\textwidth]{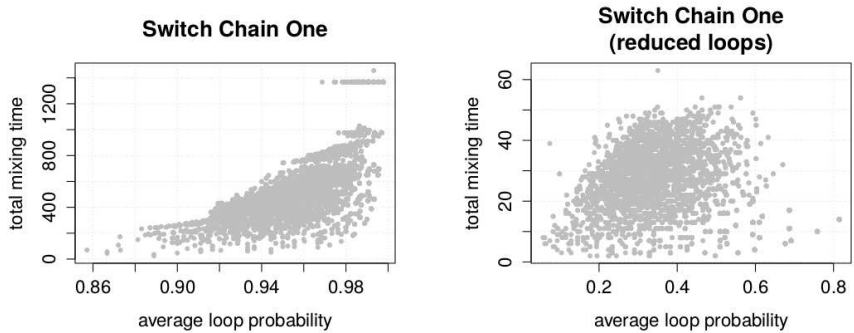}
\caption{\textbf{Influence of the average loop probability on the total mixing time.} Each data point represents a state graph, for each input instance of Switch Chain One, before and after loop reduction. }
\label{Fig:loop_avg_vs_mixing}
\end{center}
\end{figure}

\begin{figure}[h!]
\begin{center}
\includegraphics[width=\textwidth]{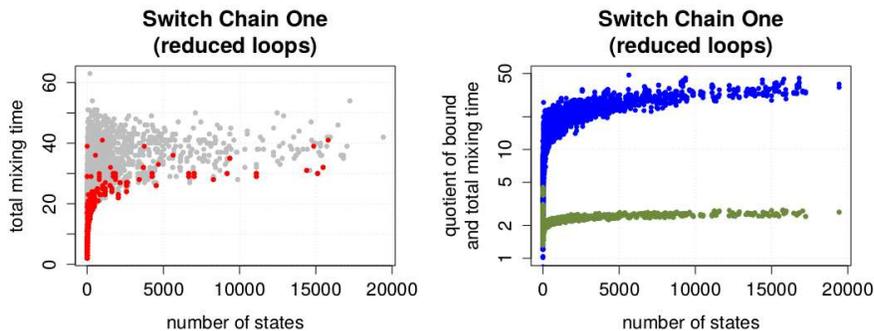}
\caption{\textbf{Properties of reduced state graphs.} (A) Size of a state graph versus its total mixing time. Regular and half-regular sequence pairs are highlighted red. (B) Quotient of congestion bound (blue) and upper spectral bound (green) with corresponding total mixing time.}
\label{Fig:loop_size_vs_mixing}
\end{center}
\end{figure}

\begin{figure}[h!]
\begin{center}
\includegraphics[width=\textwidth]{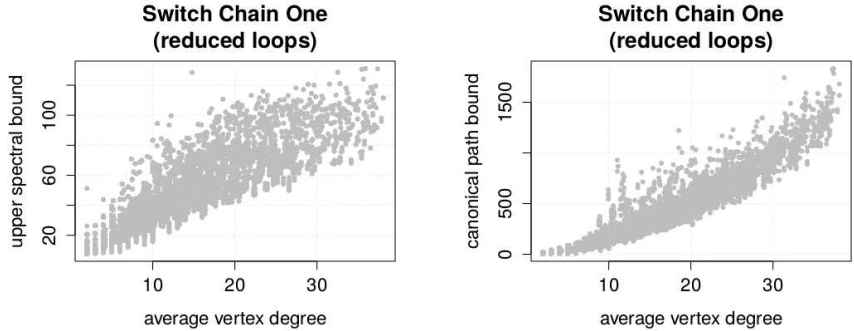}
\caption{\textbf{Influence of the average vertex degree.} Connection between average vertex degree of a state graph and its total mixing time, respectively canonical path bound.}
\label{Fig:loop_avgdeg_vs_mixing}
\end{center}
\end{figure}

We first observed that the modification indeed had a strong effect on the total mixing time (see Fig.~\ref{Fig:loop_avg_vs_mixing}).
While the largest total mixing time in our instance set is about 1400, in the set of reduced state graphs, it is 63. In addition, we observed that the total mixing time after the reduction steps no longer correlates with the average loop probability. So, the effects we observe now are more or less detached from the influence of the loops. We observed, that the effect we observed in the enumeration experiment still remain (see Fig.\ref{Fig:loop_size_vs_mixing}). The quality of the congestion bound is still bad while the largest total mixing time occurs at relatively small state graphs. 
We finished our experiments with the observation that the upper spectral bound as well as the congestion bound heavily depend on the average vertex degree of the state graph (see Fig.~\ref{Fig:loop_avgdeg_vs_mixing}). We concluded that the average degree is a structural property of a state graph which has a stronger influence on the total mixing time than its size.

\section*{Results and Discussion}

We introduced the \textsl{marathon} library and demonstrated how it can be used to support the analysis of sampling algorithms by a set of experiments.
We briefly summarise the most important observations from our experiments:
\begin{itemize}
\item The congestion bound is much larger than the total mixing time in almost all cases. The quality of the congestion bounds deteriorates with growing input size. This observation is made in reduced and non-reduced state graphs.
\item The upper spectral bound keeps close to total mixing time even when the input size grows, although it can be observed, that its quality drops slowly. Again, this observation is made in reduced and non-reduced state graphs.
\item The lower spectral bound seems to have an almost linear relationship with the total mixing time. This way, it can be used to precisely predict the total mixing time. The search for an explanation of this effect is likely a good subject for further theoretical studies.
\item Maybe the most surprising observation made in our experiments is that large total mixing time tends to occur at small state graphs.
\item The average vertex degree of a state graph has a strong relationship on its total mixing time. This observation should be investigated further. Figuring out which sequences have a tendency towards more diverse vertex degrees in their state graphs could lead to further interesting rapidly mixing graph classes.
\end{itemize}
Our observations indicate that the theoretical bound gained by the canonical path method is likely too pessimistic. Moreover, although we know the exact structure of a state graph in our experiments, which can never be the case in a normal practical scenario, the congestion bound is too large. This leads us to the conclusion that the canonical path method will never yield applicable values. Furthermore, this gives hope to the hypothesis, that the true total mixing time is smaller than existing theory is able to prove. So, we are optimistic about the applicability of the switch chains. We think that a promising next step could be to focus on two fields which need one another: investigating and the proving of special structures for state graphs depending on the problem and figuring out new relationships between structures of graphs and its eigenvalues.  

\subsection*{Future Work}

One of our next steps is to use our new tool very extensively. We want to investigate further chains and test further hypotheses. 
Moreover, we plan to add further functionality to the \textsl{marathon} library to enable additional research questions. For example, we want to compute the distribution of the mixing time. While the total mixing time  displays the maximal number of steps necessary to drop below a total variation distance of $\epsilon$, the mixing time as a function of the starting state would lead to further insights, such as average mixing time. Another additional feature would be the computation of a multi commodity flow congestion scheme, which is a generalization of the canonical path method. 
Furthermore, we want to improve the performance of existing methods. In particular, we want to add multi GPU support as well as a parallel method for the construction of the state graph. Another topic would be the parallel evaluation of the canonical path congestion method using GPUs.

\section*{Acknowledgments}
We would like to thank Matthias Müller-Hannemann for his remarks and suggestions, as well as for the fruitful discussions.


%
%
%




\begin{thebibliography}{10}

\bibitem{Kashtan2005}
Kashtan N, Itzkovitz S, Milo R, Alon U. {mfinder}: Network motif detection
  tool; 2005.
\newblock Available from: \url{http://www.weizmann.ac.il/mcb/UriAlon/}.

\bibitem{gotelliNicholas}
Gotelli NJ.
\newblock {Null Model Analysis of Species Co-Occurrence Patterns}.
\newblock Ecology. 2000;81(9):2606--2621.
\newblock Available from: \url{http://www.jstor.org/stable/177478}.

\bibitem{Jerrum1986169}
Jerrum MR, Valiant LG, Vazirani VV.
\newblock Random generation of combinatorial structures from a uniform
  distribution.
\newblock Theoretical Computer Science. 1986;43:169 -- 188.
\newblock Available from:
  \url{http://www.sciencedirect.com/science/article/pii/030439758690174X}.

\bibitem{diaconis2009markov}
Diaconis P.
\newblock The markov chain monte carlo revolution.
\newblock Bulletin of the American Mathematical Society. 2009;46(2):179--205.

\bibitem{Sinclair92}
Sinclair A.
\newblock {Improved bounds for mixing rates of Markov chains and multicommodity
  flow}.
\newblock In: Simon I, editor. LATIN '92. vol. 583 of Lecture Notes in Computer
  Science. Springer Berlin Heidelberg; 1992. p. 474--487.
\newblock Available from: \url{http://dx.doi.org/10.1007/BFb0023849}.

\bibitem{SinclairLectures}
Sinclair A. C{S}294: Markov Chain Monte Carlo: Foundations \& Applications;
  2009.
\newblock Last visited on March 2014.
\newblock
  \url{http://www.cs.berkeley.edu/}\url{\~}\url{sinclair/cs294/f09.html}.

\bibitem{LevinPeresWilmer2006}
Levin DA, Peres Y, Wilmer EL.
\newblock {Markov chains and mixing times}.
\newblock American Mathematical Society; 2006.

\bibitem{Lovasz93randomwalks}
Lov\'asz L.
\newblock Random Walks on Graphs: A Survey.
\newblock In: {Mikl\'os} D, {S\'os} VT, {Sz\H{o}nyi} T, editors. Combinatorics,
  Paul Erd\H{o}s is Eighty. vol.~2. Budapest: J\'anos Bolyai Mathematical
  Society; 1996. p. 353--398.

\bibitem{cudaToolkit}
Nvidia.
\newblock CUDA Toolkit; 2015.
\newblock Available from: \url{https://developer.nvidia.com/cuda-toolkit}.

\bibitem{OpenMP}
Dagum L, Menon R.
\newblock OpenMP: An Industry-Standard API for Shared-Memory Programming.
\newblock IEEE Comput Sci Eng. 1998 Jan;5(1):46--55.
\newblock Available from: \url{http://dx.doi.org/10.1109/99.660313}.

\bibitem{openblas}
Xianyi Z.
\newblock openBLAS; 2015.
\newblock Available from: \url{http://www.openblas.net/}.

\bibitem{cublas}
Nvidia.
\newblock cuBLAS; 2015.
\newblock Available from: \url{https://developer.nvidia.com/cublas}.

\bibitem{cublasxt}
Nvidia.
\newblock cuBLAS-XT; 2015.
\newblock Available from: \url{https://developer.nvidia.com/cublasxt}.

\bibitem{arpack}
Lehoucq RB, Sorensen DC, Yang C.
\newblock {Arpack User's Guide: Solution of Large-Scale Eigenvalue Problems
  With Implicityly Restorted Arnoldi Methods (Software, Environments, Tools)}.
\newblock Soc for Industrial \& Applied Math;.
\newblock Available from: \url{http://www.worldcat.org/isbn/0898714079}.

\bibitem{broder86}
Broder AZ.
\newblock {How hard is it to marry at random? (On the approximation of the
  permanent)}.
\newblock In: Proceedings of the Eighteenth Annual ACM Symposium on Theory of
  Computing. STOC '86. New York, NY, USA: ACM; 1986. p. 50--58.
\newblock Available from: \url{http://doi.acm.org/10.1145/12130.12136}.

\bibitem{Jerrum:1989:AP:76071.76077}
Jerrum M, Sinclair A.
\newblock {Approximating the permanent}.
\newblock SIAM J Comput. 1989 Dec;18(6):1149--1178.
\newblock Available from: \url{http://dx.doi.org/10.1137/0218077}.

\bibitem{Diaconis99statisticalproblems}
Diaconis P, Graham R, Holmes SP.
\newblock {Statistical problems involving permutations with restricted
  positions}.
\newblock In: IMS Lecture Notes Monogr. Ser; 1999. p. 195--222.

\bibitem{JerrumSinclairVigoda04}
Jerrum M, Sinclair A, Vigoda E.
\newblock {A polynomial-time approximation algorithm for the permanent of a
  matrix with nonnegative entries}.
\newblock Journal of the ACM. 2004;51:671--697.

\bibitem{Bezakova06acceleratingsimulated}
Bezáková I, Stefankovič D, Vazirani VV, Vigoda E.
\newblock Accelerating Simulated Annealing for the Permanent and Combinatorial
  Counting Problems.
\newblock In: In Proceedings of the 17th Annual ACM-SIAM Symposium on Discrete
  Algorithms (SODA). ACM Press; 2006. p. 900--907.

\bibitem{Tutte52}
Tutte WT.
\newblock {The factors of graphs}.
\newblock Canadian J of Mathematics. 1952;4:314--328.

\bibitem{Kannan97}
Kannan R, Tetali P, Vempala S.
\newblock Simple Markov-chain algorithms for generating bipartite graphs and
  tournaments.
\newblock Random Structures \& Algorithms. 1999;14(4):293--308.

\bibitem{Petersen1891}
Petersen J.
\newblock {Die Theorie der regulären graphs}.
\newblock Acta Mathematica. 1891;15:193--220.

\bibitem{Miklos13}
Mikl{\'o}s I, Erd{\"o}s PL, Soukup L.
\newblock {Towards Random Uniform Sampling of Bipartite Graphs with given
  Degree Sequence}.
\newblock Electr J Comb. 2013;20(1).

\bibitem{DBLP:conf/soda/Greenhill15}
Greenhill CS.
\newblock The switch Markov chain for sampling irregular graphs (Extended
  Abstract).
\newblock In: Proceedings of the Twenty-Sixth Annual {ACM-SIAM} Symposium on
  Discrete Algorithms, {SODA} 2015, San Diego, CA, USA, January 4-6, 2015;
  2015. p. 1564--1572.
\newblock Available from: \url{http://dx.doi.org/10.1137/1.9781611973730.103}.

\bibitem{Berger:2010}
Berger A, M{\"{u}ller-Hannemann} M.
\newblock {Uniform sampling of digraphs with a fixed degree sequence}.
\newblock In: Proceedings of the 36th International Conference on
  Graph-Theoretic Concepts in Computer Science. WG'10. Berlin, Heidelberg:
  Springer-Verlag; 2010. p. 220--231.
\newblock Full version available from Arxiv:0912.0685v3.

\bibitem{McKay201494}
McKay BD, Piperno A.
\newblock Practical graph isomorphism, (II).
\newblock Journal of Symbolic Computation. 2014;60(0):94 -- 112.
\newblock Available from:
  \url{http://www.sciencedirect.com/science/article/pii/S0747717113001193}.

\end{thebibliography}

\section*{Supporting Information}

\noindent
\textbf{S1 Dataset. Results of the enumeration experiment.} The results of the enumeration experiment as comma separated value files.\\

\noindent
\textbf{S2 Dataset. Results of the scaling experiment.} The results of the scaling experiment  as  comma separated value files.\\

\noindent
\textbf{S3 Dataset. Results of the loop reduction experiment.} The results of the loop reduction experiment as a comma separated value file.

\end{document}